\documentclass{emulateapj}
\usepackage{scrextend}

\usepackage[backref=section]{hyperref}

\usepackage[dvipsnames]{xcolor}
\colorlet{mylinkcolor}{Maroon}
\colorlet{mycitecolor}{MidnightBlue}
\colorlet{myurlcolor}{MidnightBlue}
\hypersetup{ 
	colorlinks   = true,
	citecolor    = mycitecolor,
	linkcolor    = mylinkcolor,
	urlcolor	= myurlcolor
}

\usepackage{etoolbox}
\makeatletter
\patchcmd{\BR@backref}{\newblock}{\newblock[}{}{}
\patchcmd{\BR@backref}{\par}{]\par}{}{}
\makeatother
\usepackage[all]{hypcap}

\usepackage{graphicx}
\usepackage{epsfig}
\usepackage{natbib}
\usepackage{multirow}
\usepackage{amsmath,amssymb}
\usepackage{rotating}
\usepackage{pdfpages}
\usepackage{xcolor}
\usepackage[flushleft]{threeparttable}
\usepackage{booktabs}

\usepackage{CJKutf8}
\usepackage{ucs}
\usepackage[encapsulated]{CJK}
\newcommand{\myfont}{bsmi} 

\submitted{ApJ in press}

\lefthead{Chen et al.}
\righthead{{\sc The clustering of faint SMGs}}

\begin{document}


\title{{Faint Submillimeter Galaxies identified through their optical/near-infrared colours I: spatial clustering and halo masses}}

\begin{CJK}{UTF8}{\myfont} 
\author{Chian-Chou Chen (陳建州)\altaffilmark{1,2}, Ian Smail\altaffilmark{1,2}, A. M. Swinbank\altaffilmark{1,2}, James M. Simpson\altaffilmark{3}, Omar Almaini\altaffilmark{4}, Christopher J. Conselice\altaffilmark{4}, Will G. Hartley\altaffilmark{4,5}, Alice Mortlock\altaffilmark{4,3}, Chris Simpson\altaffilmark{6}, Aaron Wilkinson\altaffilmark{3}}

\email{chian-chou.chen@durham.ac.uk}
\altaffiltext{1}{Centre for Extragalactic Astronomy, Department of Physics, Durham University, South Road, Durham DH1 3LE, UK}
\altaffiltext{2}{Institute for Computational Cosmology, Durham University, South Road, Durham DH1 3LE, UK}
\altaffiltext{3}{Institute for Astronomy, University of Edinburgh, Royal Observatory, Blackford Hill, Edinburgh EH9 3HJ, UK}
\altaffiltext{4}{University of Nottingham, School of Physics and Astronomy, Nottingham, NG7 2RD UK}
\altaffiltext{5}{ETH Z\"{u}rich, Institut f\"{u}r Astronomie, HIT J 11.3, Wolfgang-Pauli-Strasse 27, CH-8093 Z\"{u}rich, Switzerland}
\altaffiltext{6}{Astrophysics Research Institute, Liverpool John Moores Uni- versity, Liverpool Science Park, 146 Brownlow Hill, Liverpool L3 5RF}

\begin{abstract}
The properties of submillimeter galaxies (SMGs) that are fainter than the confusion limit of blank-field single-dish surveys ($S_{850} \lesssim$ 2\,mJy) are poorly constrained. Using a newly developed color selection technique, Optical-Infrared Triple Color (OIRTC), that has been shown to successfully {select} such faint SMGs, we identify a sample of 2938 OIRTC-selected galaxies, dubbed Triple Color Galaxies (TCGs), in the UKIDSS-UDS field. We show that these galaxies have a median 850\,$\mu$m flux of S$_{850} = 0.96\pm0.04$\,mJy (equivalent to a star-formation rate SFR $\sim60-100$\,M$_\odot$ yr$^{-1}$ {based on SED fitting}), representing the first large sample of faint SMGs that bridges the gap between bright SMGs and normal star-forming galaxies {in S$_{850}$ and $L_{\rm IR}$}. We assess the basic properties of TCGs and their relationship with other galaxy populations at $z\sim2$. We measure the two-point autocorrelation function for this population and derive a typical halo mass of log$_{10}$(M$_{\rm halo}$) $=12.9^{+0.2}_{-0.3}$, $12.7^{+0.1}_{-0.2}$, and $12.9^{+0.2}_{-0.3}$\,$h^{-1}$M$_\odot$ at $z=1-2$, $2-3$, and $3-5$, respectively. Together with the bright SMGs (S$_{850} \gtrsim 2$\,mJy) and a comparison sample of less far-infrared luminous star-forming galaxies, we find a lack of dependence between spatial clustering and S$_{850}$ (or SFR), suggesting that the difference between these populations may lie in their local galactic environment. Lastly, on the scale of $\sim8-17$\,kpc at $1<z<5$ we find a tentative enhancement of the clustering of TCGs over the comparison star-forming galaxies, suggesting that some faint SMGs are physically associated pairs, perhaps reflecting a merging origin in their triggering.
 
\end{abstract}


\section{Introduction}\label{sec:intro}
The development of far-infrared(FIR)/submillimeter single-dish surveys has successfully resolved a moderate fraction of the cosmic FIR/submillimeter background light into bright, rare, very dusty sources that host the most intense star formation across the Universe. Panchromatic follow-up studies have revealed that these dusty galaxies are mostly located at $z\sim1-3$ (e.g., \citealt{Chapman:2005p5778,Simpson:2014aa,Chen:2016aa}), and they have a total infrared luminosity ($L_{\rm IR}$) that is comparable to the local Ultra-luminous infrared galaxies (ULIRGs) with $L_{\rm IR}$ greater than few times\,10$^{12}$\,L$_{\odot}$ and sometime reaching over 10$^{13}$\,L$_{\odot}$. While many of their basic physical properties such as number counts, stellar masses and the triggering mechanism of star formation, are still under debate \citep{Karim:2013fk,Chen:2013gq,Michaowski:2014ab,Targett:2013aa,Chen:2015aa}, these submillimeter galaxies (SMGs; \citealt{Smail:1997p6820}) or dusty star-forming galaxies (DSFGs; \citealt{Casey:2014aa}) are nevertheless excellent laboratories for testing star-formation laws in extreme galactic environments (e.g., \citealt{Hodge:2015aa}), as well as models of galaxy formation and evolution in general (e.g., \citealt{Lacey:2015aa}).

The main challenge for exploiting large-scale ($\sim$degree$^2$) FIR/submillimeter surveys to study the SMG population has been the fact that these are generally diffraction-limited observations. At 850\,$\mu$m, the highest angular resolution currently achieved by the SCUBA-2 camera mounted on the 15-meter James Clerk Maxwell Telescope (JCMT) is 15$''$. This modest resolution has resulted in slow progress in understanding SMGs, {which is largely due to two factors}: First, coarse resolution means it is not straightforward to identify the correct counterparts at other wavelengths, leading to contaminated results regarding the physical properties of SMGs.
Interferometric follow-up observations can eventually pinpoint the SMG positions to sub-arcsecond accuracy (e.g., \citealt{Gear:2000aa,Younger:2007p6982, Barger:2012lr,Hodge:2013lr,Simpson:2015ab}), however this means an extra step and thus observationally expensive to match the counterparts. 

Secondly, and perhaps more importantly, the amount of source blending caused by the poor resolution increases with increased image depth, until the point the map is completely covered, meaning all the signal in the map is contributed by real astronomical sources. This point of saturation, or ``confusion limit'' \citep{Condon:1974qy, Hogg:2001uq}, prevents the detection of faint sources regardless of the amount of exposure. In blank-field surveys, the confusion limit is $\sim$20\,mJy at 250, 350, and 500\,$\mu$m \citep{Nguyen:2010kx} based on the data taken by the {\it Herschel} Space Observatory (\citealt{Pilbratt:2010lr}; hereafter {\it Herschel}), and $\sim$2\,mJy at 850\,$\mu$m for the JCMT (e.g., \citealt{Coppin:2006p9123}).

The fundamental impact of confusion is that blank-field submillmeter single-dish surveys can not directly detect faint SMGs that are below this limit. As a result, faint SMGs (S$_{850} \lesssim 2$\,mJy) still remain poorly understood, both observationally and hence also theoretically, despite the fact that cosmologically they contribute $\sim$80\% of the {850\,$\mu$m} extragalactic background light (EBL; e.g., \citealt{Cowie:2002p2075}). In addition, from the galaxy formation and evolution point of view, faint SMGs with 850\,$\mu$m flux of $S_{850}\sim0.4-2$\,mJy are expected to have a total infrared luminosity similar to the local LIRGs (10$^{11} \leq L_{\rm IR} \leq $\,10$^{12}$\,L$_{\odot}$) and thus represent a key population that {bridges the gap in $L_{\rm IR}$ between} the violent star-forming galaxies such as bright SMGs and normal star-forming galaxies such as Lyman Break Galaxies (LBGs) or $BzK$ galaxies, holding critical information about a potentially important transitional period of galaxy evolution.

Techniques have been developed to study faint SMGs. First, by conducting surveys in the field of galaxy clusters, any background faint SMGs can be magnified to a detectable flux level through strong gravitational lensing. Samples of faint SMGs have been discovered  and the submillimeter number counts have been constructed using this technique \citep{Smail:1997p6820,Cowie:2002p2075,Knudsen:2008p3824,Johansson:2011zr,Chen:2013fk,Chen:2013gq,Hsu:2016aa}. The drawback of this method, however, is that the intrinsic properties of individual faint SMG sometimes suffer large uncertainties due to the systematics of the lensing models and the uncertainty of the source redshift, in particular for strongly lensed sources (e.g., \citealt{Chen:2011p11605}).

With the advent of ALMA, the second approach to detecting faint SMGs have started to emerge. Deep ALMA imaging taken for the primary target of interests are sometimes deep enough to make serendipitous faint SMG detections \citep{Hatsukade:2013aa,Hodge:2013lr,Ono:2014aa,Carniani:2015aa,Fujimoto:2016aa,Oteo:2015aa,Simpson:2015ab}. However, so far such studies have only surveyed relatively small areas and the clustering properties of faint SMGs are neither well constrained nor unbiased \citep{Ono:2014aa, Fujimoto:2016aa}. 

Finally, conducting blank-field ALMA mosaic observations could offer an unbiased approach in detecting faint SMGs \citep{Kohno:2016aa,Dunlop:2016aa}, however, surveying degree$^2$ scale area with ALMA will remain challenging and time consuming even in its full capability. Therefore, an efficient way in finding faint SMGs across large areas, in particular using multi-wavelength data readily available in extragalactic legacy fields, would provide an opportunity to address fundamental questions such as what is the clustering strength of the faint SMGs?

The spatial clustering strength, as measured from the two-point autocorrelation functions, can provide important information about the relationship between different galaxy populations (for a review see \citealt{Cooray:2002aa}). Under the standard $\Lambda$CDM cosmology, galaxies that experience the same evolutionary track should reside in halos with similar masses at any given redshift. Therefore by measuring the halo mass, which can be inferred from the spatial clustering strength, we can test links between various galaxy populations and galaxy evolution models.

In this paper, we propose a new method to select faint SMGs in wide fields, {which exploits our recent findings of the distinct optical and infrared color space that the faint SMGs occupy}. In \citet{Chen:2016aa}, we developed a new color selection technique, Optical-Infrared Triple Color (OIRTC), using ($z-K$), ($K-$[3.6]) and ([3.6]-[4.5]) to select the SMG counterpart candidates based on a training set of SMGs from an ALMA pilot study of a subset of the bright SCUBA-2 sample in the UKIDSS-UDS field. Using this selection we found that 87\% of the OIRTC-selected galaxies are confirmed as submillimeter sources by ALMA at a 850\,$\mu$m detection limit of $\sim$1\,mJy. This accuracy is as good as that of the traditional corrected-Poissonian probability identification technique ($p$-value) using radio counterparts (e.g., \citealt{Ivison:2002uq}). The advantage of the OIRTC selection, however, is that it does not need the single-dish detection as a prior to find the SMGs, and it can simply be applied to any samples as long as there are appropriate photometric observations. 


\begin{figure*}
	\begin{center}
		\leavevmode
		\includegraphics[scale=1]{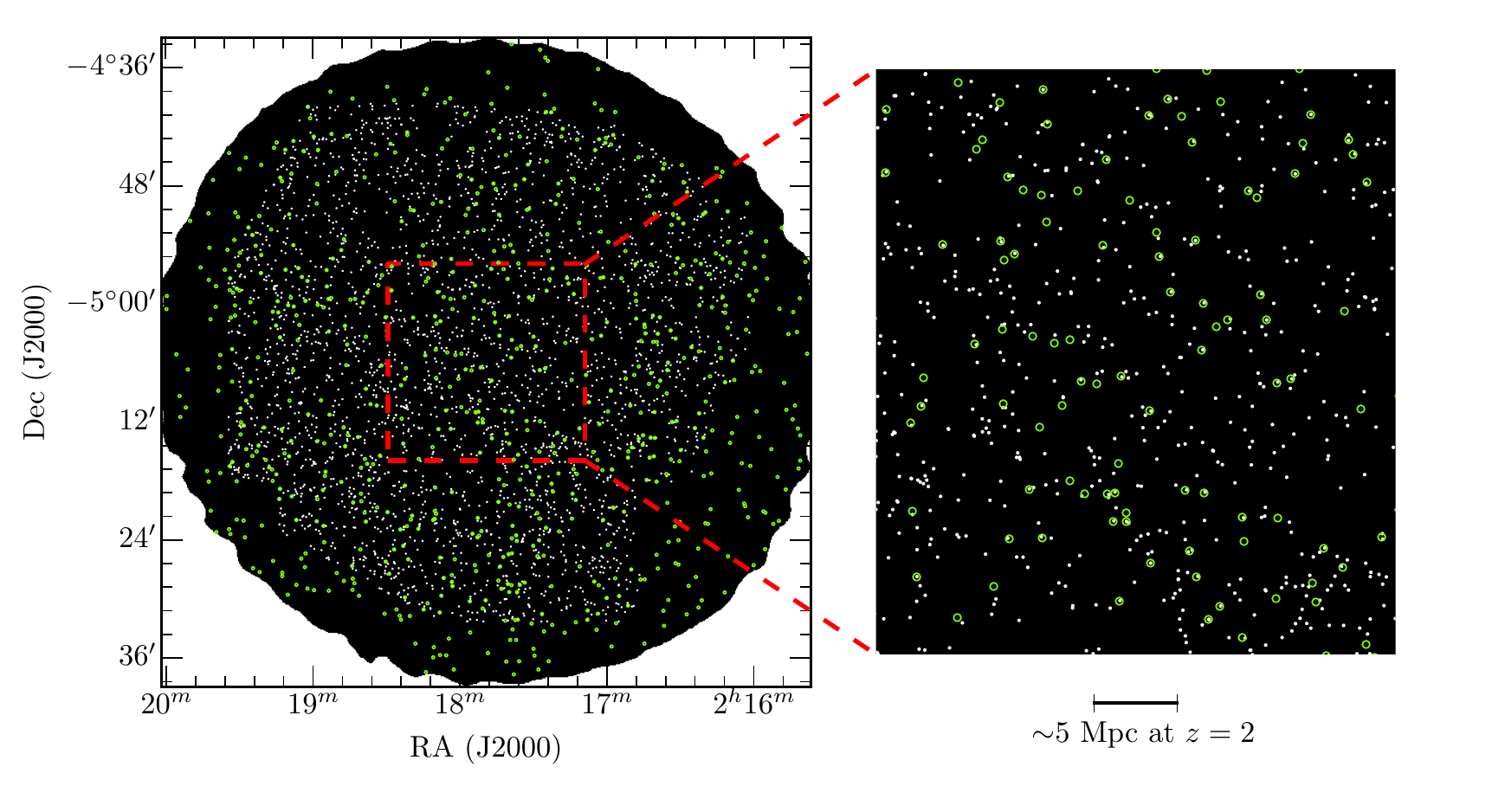}
		\caption{An overview of UKIDSS-UDS field, with the black background representing the SCUBA-2 footprint (Geach et al. 2016). TCGs are marked as white small dots and the SCUBA-2 $\geq4$\,$\sigma$ detections are plotted as green circles. All TCGs and comparison samples are covered by the SCUBA-2 850\,$\mu$m imaging, as well as by the SPIRE imaging at 250, 350, and 500\,$\mu$m taken by {\it Herschel} \citep{Oliver:2010p11204, Swinbank:2014ul}.
		}
		\label{overview}
	\end{center}
\end{figure*}

In this paper, we describe the basic properties of the 2938 OIRTC-selected galaxies (hereafter TCGs) located in the UDS field in terms of 850\,$\mu$m flux, redshift, stellar mass, and rest-frame colors in the optical and near-infrared, to put them into the context of the general galaxy population. With almost 3000 sources across a single $\sim$1 degree$^2$ field, we end this paper by presenting a clustering {analysis}, which yields the first measurement on the halo mass of the faint SMGs. In a subsequent paper we will investigate the stellar morphology and sizes for TCGs that are located in the CANDELS field.

This paper is based on the multi-wavelength data taken in the UKIDSS-UDS field, for which the details are described in \autoref{sec:obs}. The selections of the TCGs, as well as comparison samples of less active star-forming galaxy and quiescent galaxy that are matched to the TCGs in stellar mass and redshift, are also described in \autoref{sec:obs}. The basic properties of the TCGs and the implication on the selection of the high-redshift quiescent galaxies are shown in \autoref{sec:tcg}. We present the results of the clustering analyses in \autoref{sec:clustering}, discussion in \autoref{sec:discussion}. Finally the summary is given in \autoref{sec:sum}. Throughout this paper we adopt the AB magnitude system \citep{Oke:1983aa}, and we assume the {\it Planck} cosmology: H$_0 =$\,67.77\,km\,s$^{-1}$ Mpc$^{-1}$, $\Omega_M = $\,0.31, and $\Omega_\Lambda =$\,0.69 \citep{Planck-Collaboration:2014aa}.

\section{Data}\label{sec:obs}
\subsection{Parent sample, ancillary data, photometric redshifts, and stellar masses}\label{sec:parent}
Our sample is drawn from the $K$-band image of the UKIRT Infrared Deep Sky Survey (UKIDSS; \citealt{Lawrence:2007aa}) data release 8 (DR8). The Ultra Deep Survey (UDS) field is the deepest of the five sub-surveys of UKIDSS, consisting of four Wide-Field Camera (WFCAM; \citealt{Casali:2007aa}) pointings, covering 0.77 square degrees in $J$, $H$ and $K$ bands. The DR8 release contains all UDS data taken from 2005 to 2010. The 5\,$\sigma$ median depths are $J = 24.9$, $H = 24.2$ and $K = 24.6$ (in a 2$''$ diameter aperture). The parent catalogue was extracted using {\sc sextractor} \citep{Bertin:1996zr} on the deep $K$-band image, and the extraction parameters were designed to recover both point-like and extended low-surface-brightness sources. Detailed descriptions of mosaicing, catalogue extraction and depth estimation will be presented in Almaini et al. (in preparation). After selecting the $\geq5$\,$\sigma$ $K$-band detections in a 2$''$ diameter aperture, masking bad regions, removing bright stars and image artefacts produced by amplifier cross-talk, a parent sample of 115,671 sources across 0.6 degree$^2$ was constructed for our analysis. To exploit the rich ancillary data in this field, we only consider sources that have full multi-wavelength data coverage from UV to infrared (see below).  

In addition, the UDS field was covered by the Megacam $u'$-band on the Canada-France-Hawaii Telescope (CFHT), with a 5\,$\sigma$ depth reaching $u'=$\,26.75 in a 2$''$ diameter aperture. The field was also observed by the Subaru telescope using Suprime-Cam in five broadband filters, $B$, $V$, $R_c$, $i'$, and $z'$, to the limiting depths of $B =$\,28.4, $V =$\,27.8, $R_c =$\,27.7, $i' =$\,27.7, and $z' =$\,26.6, respectively (3\,$\sigma$, 2$''$ diameter apertures). Details of the Suprime-Cam survey are provided in \citet{Furusawa:2008aa}. The mid-infrared IRAC data were obtained by the {\it Spitzer} Legacy Program SpUDS (PI: Dunlop), reaching 5\,$\sigma$ depths of 24.2 and 24.0 AB magnitude at 3.6 and 4.5\,$\mu$m. The UDS field is also observed in FIR by {\it Herschel} with the SPIRE instrument at 250, 350, and 500\,$\mu$m, {which were taken as part of the  Herschel Multi-tiered Extragalactic Survey (HerMES; \citealt{Oliver:2010p11204}).} In the submillimeter, the UDS field is uniformly covered with SCUBA-2 camera at 850\,$\mu$m as part of the SCUBA-2 Cosmology Legacy Survey (S2CLS; Geach et al. 2016; \autoref{overview}). There are 716 SCUBA-2 sources detected at $\geq4$\,$\sigma$ in UDS with 850\,$\mu$m fluxes of S$_{850} \gtrsim 3$\,mJy, and multi-wavelength identifications are presented in \citet{Chen:2016aa}. Finally, X-ray data were obtained as part of the Subaru-{\it XMM/Newton} Deep Survey (SXDS), consisting of seven contiguous fields with a total exposure of 400\,ks in the 0.2--10 keV band \citep{Ueda:2008aa}. 

Eleven-band photometry ($UBVRIzJHK$[3.6][4.5]) was measured with 3$''$ diameter apertures placed on each aligned image at the position of the $K$-band sources, motivated by the fact that $K$-band is generally a good stellar mass indicator at moderate redshifts and moreover is less affected by dust compared to other optical/NIR bands, with a data quality that is deeper and has a higher angular resolution compared to that of the IRAC bands. To account for the correlated noise that is not represented in the weight maps, the magnitude uncertainties estimated by {\sc sextractor} are corrected by scaling the weight maps such that the uncertainty in source-free regions matches the rms measured from apertures placed on the science image. Three of the bands (the CFHT $u'$ band and the two IRAC channels) required aperture corrections to their photometry in order to obtain correct colors. This correction was performed based on smoothing the $K$-band images to the appropriate PSF and re-computing the aperture photometry to evaluate the expected changes. More details can be found in \citet{Hartley:2013aa}.

Photometric redshifts ($z_{\rm photo}$) have been derived for the DR8 parent sample, and the full description can be found in \citet{Hartley:2013aa}, \citet{Mortlock:2013aa} and \citet{Mortlock:2015aa}. In summary, the photometric 
redshifts are estimated using the {\sc eazy} template-fitting package \citep{Brammer:2008aa} through a maximum likelihood analysis. The default set of six templates does not sufficiently represent all of the galaxies, in particular the $u'$-band flux is significantly overestimated for the blue sources at high redshift. A seventh template is therefore constructed by applying a small amount of Small Magellanic Cloud-like extinction \citep{Prevot:1984aa} to the bluest template in {\sc eazy}. The accuracy of the photometric redshift was assessed by comparing to the existing spectroscopic redshifts in the UDS. A large fraction of these $z_{spec}$ came from the UDSz, a European Southern Observatory large spectroscopic survey (ID:180.A-0776; Almaini et al., in preparation) and also from the literature (see \citealt{Simpson:2012ab} and references therein). After excluding bright X-ray and radio sources that are likely to be AGNs \citep{Simpson:2006aa, Ueda:2008aa}, we found a dispersion in $(z_{\rm photo}-z_{\rm spec})/(1+z_{\rm spec})$, after excluding outliers ($\Delta z/(1+z_{\rm spec}) > 0.15$; $<$\,4\%), is $\Delta z/(1+z_{\rm spec}) \sim$\,0.031 \citep{Hartley:2013aa}.

\begin{figure}
	\begin{center}
		\leavevmode
		\includegraphics[scale=0.49]{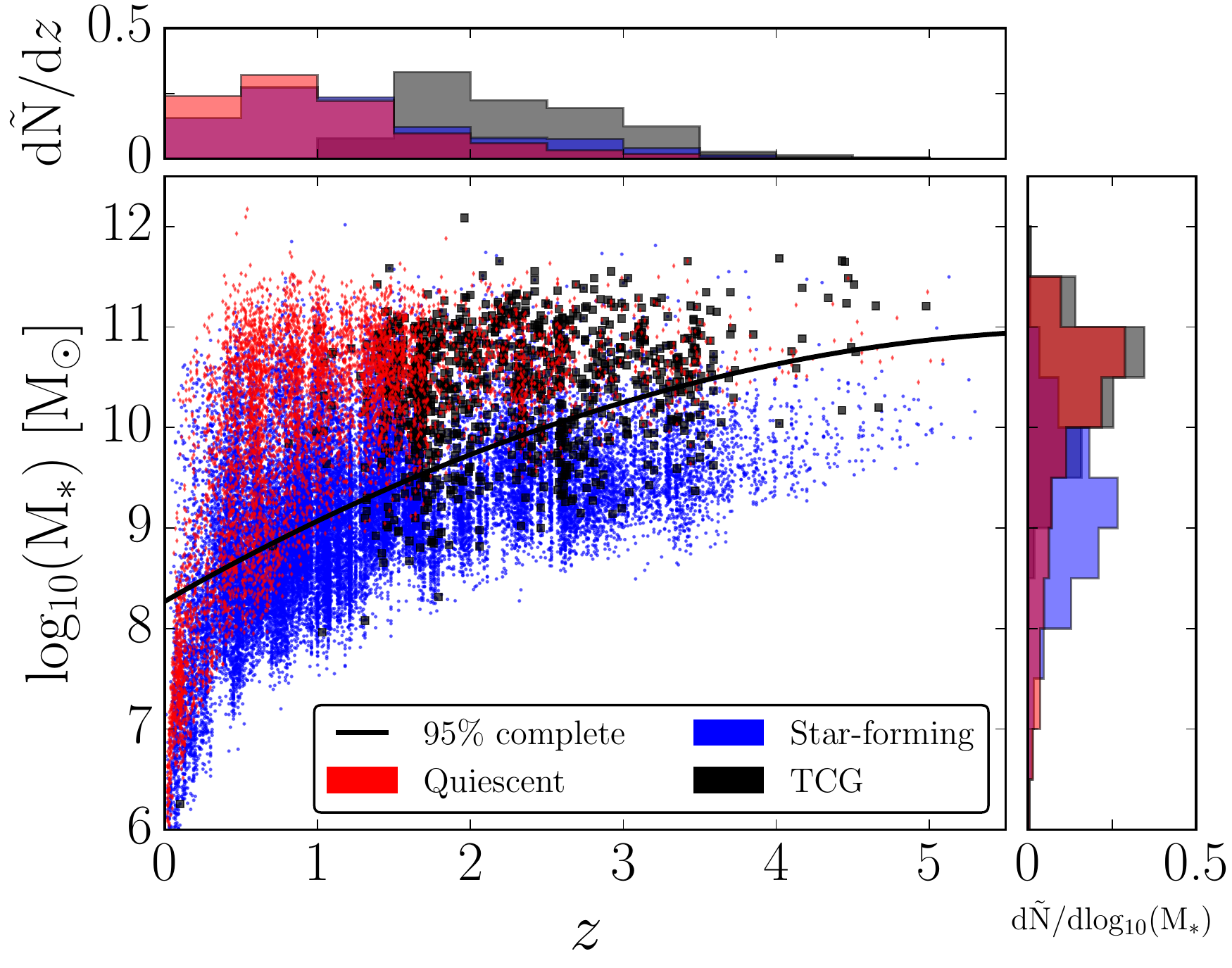}
		\caption{Estimated stellar mass versus redshift for our parent $K$-band selected sample in the UKIDSS-UDS field, color coded based on classifications of quiescent galaxies, star-forming galaxies, and TCGs. To avoid overcrowding we randomly draw a third of the sample for each population. The two histograms are normalized to the total number of sources in each galaxy category. TCGs are massive and high-redshift galaxies. The black curve shows the 95\% completeness estimated by \citet{Hartley:2013aa}.
		}
		\label{massz}
	\end{center}
\end{figure}

The stellar masses and the rest-frame luminosities were derived by using a multicolor stellar population fitting technique. As explained in detail in \citet{Hartley:2013aa} and \citet{Mortlock:2013aa}, the eleven-band photometry ($UBVRIzJHK$[3.6][4.5]) were fit to a large grid of synthetic spectral energy distributions (SEDs) constructed from the stellar population models of \citet{Bruzual:2003aa}, assuming a Chabrier initial mass function \citep{Chabrier:2003aa}. An exponentially declining star-formation history is assumed and characterised with an e-folding time, various ages, metallicities and dust extinctions. The 95 percent mass completeness as a function of redshift is estimated following \citet{Pozzetti:2010aa}, and can be described with a polynomial function $M_{\textrm lim} = 8.27+0.87z-0.07z^2$, which is shown in \autoref{massz}.

\subsection{TCG selection}
Our selection is trained on the identifications from an ALMA pilot study of a subset of the brighter SCUBA-2 sources in UDS. In \citet{Chen:2016aa} we found that near-infrared detected SMGs occupy a relatively well-defined region in the ($z-K$), ($K-$[3.6]), and ([3.6]-[4.5]) color space, in a sense that they appear to be red in all three colors. We quantitatively derived the selection limits by weighted averaging the fractional number density distribution ($\langle f_{\rm OIRTC}\rangle$) obtained in each color for both SMGs and less strongly star-forming galaxies, and we found a cut of $\langle f_{\rm OIRTC}\rangle \geq 0.05$ that best selects the SMG candidates in terms of high completeness and low contamination.

\begin{figure}
	\begin{center}
		\leavevmode
		\includegraphics[scale=0.48]{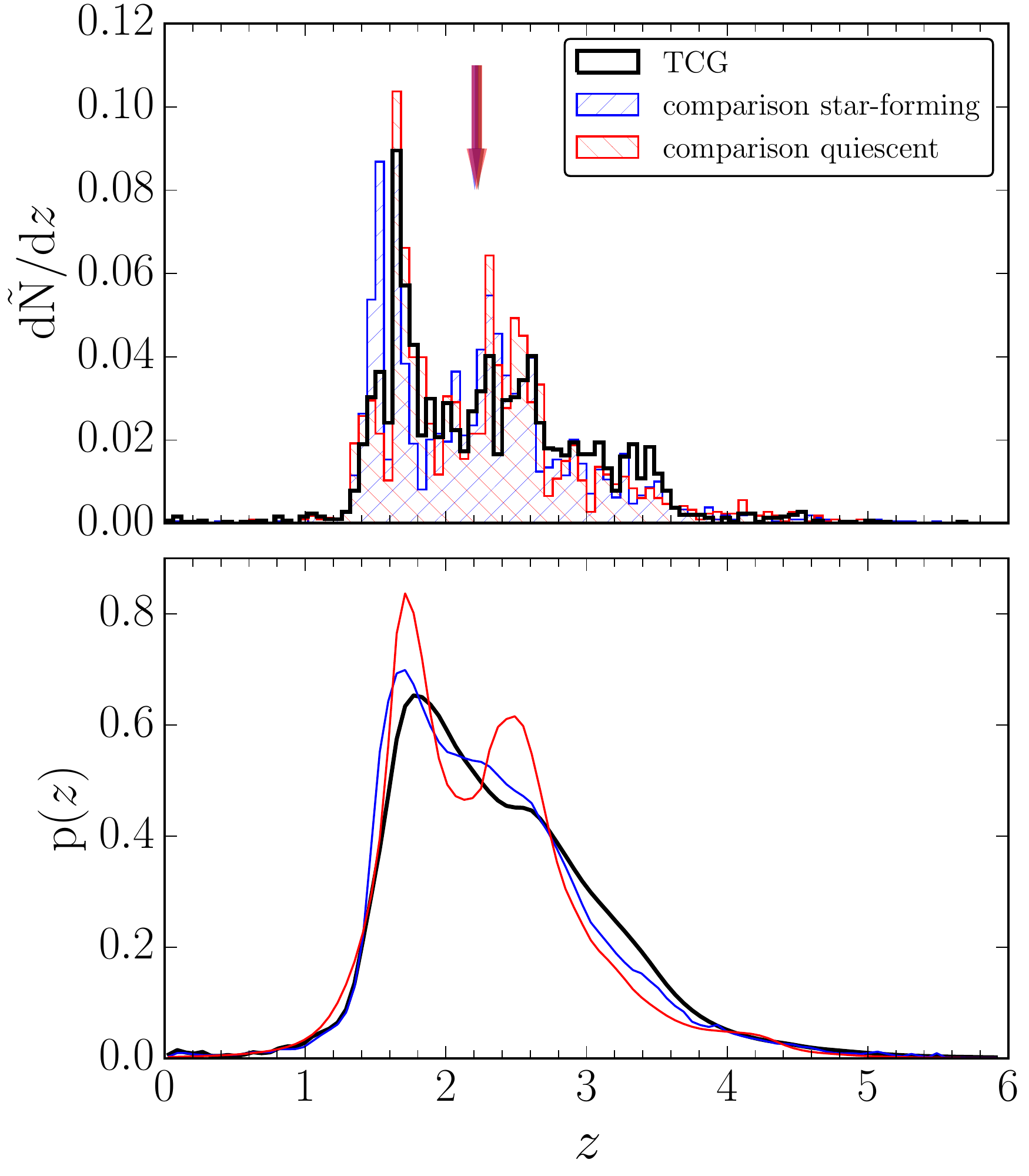}
		\caption{{\it Top}: Normalized histogram of photometric redshift for TCGs and comparison samples of quiescent galaxies and less FIR luminous star-forming galaxies. The median values are marked by the downward arrows in the corresponding colors above the histogram, with the width representing the bootstrapped uncertainties. All three populations are matched in redshifts as all three arrows overlap. {\it Bottom}: The probability distribution of redshift for all three samples based on the SED fitting results.}
		\label{zhist}
	\end{center}
\end{figure}

In practice, based on the training process presented in \citet{Chen:2016aa}, in our $K$-band parent sample described in \autoref{sec:parent} we only consider sources that have at least two measurements from our three colors ($\geq$3\,$\sigma$ in both wavebands used in each color). For each source that satisfies the criteria, we use the $f_{\rm OIRTC}$ model provided in the Table 1 of \citet{Chen:2016aa} and compute $\langle f_{\rm OIRTC}\rangle$ based on the Equation 2 in \citet{Chen:2016aa}.

From the $K$-band parent sample, we select a total of 2938 TCGs using the OIRTC technique. This TCG sample is our main science sample, and the sky positions of these TCGs are plotted on the SCUBA-2 footprint in \autoref{overview}. All TCGs are covered by the SCUBA-2 map with deep 850\,$\mu$m data. We also plot in \autoref{massz} the stellar mass-redshift distribution of TCGs, along with that of the rest of the parent sample, color-coded separately for the star-forming galaxies and quiescent galaxies based on the rest-frame $UVJ$ color cuts proposed by \citet{Williams:2009aa}. Due to the requirement of red colors in ($z-K$), which inherently selects galaxies that have strong breaks (Balmer or 4000$\AA$) or highly reddened restframe UV SED redshifted into this color regime, TCGs are mostly located at $z>1.5$ with a median of $z=2.23\pm0.02$. In addition, TCGs are on average quite massive, with a median stellar mass of log$_{10}$(M$_\star$/M$_\odot$) = $10.51\pm0.02$.

\subsection{{Comparison} sample}
\begin{figure}
	\begin{center}
		\leavevmode
		\includegraphics[scale=0.48]{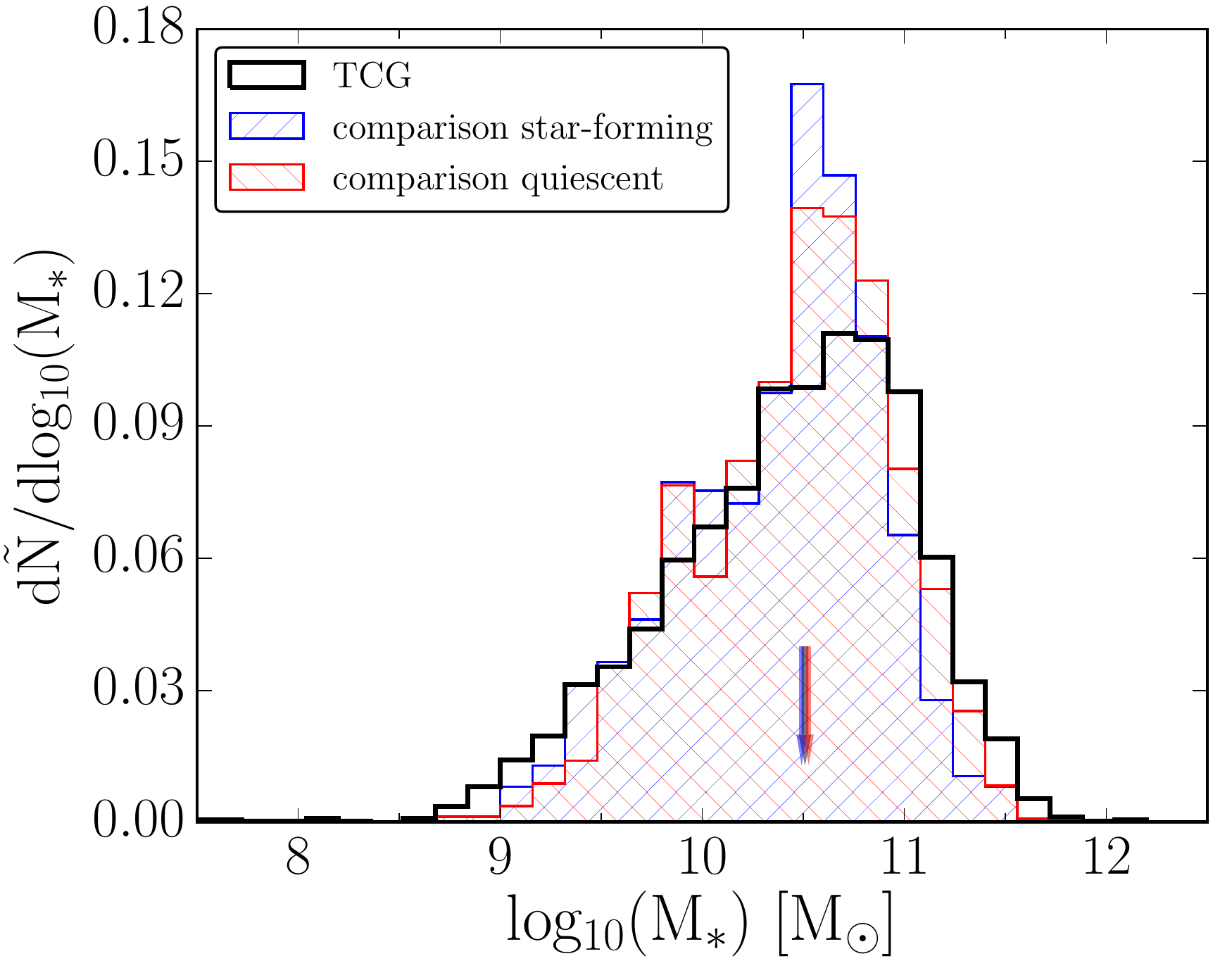}
		\caption{Normalized histogram of stellar mass for TCGs, comparison quiescent galaxies, and comparison star-forming galaxies. The median values are marked by the downward arrows in the corresponding colors above the histogram, with the width representing the bootstrapped uncertainties. All three populations are matched in stellar mass as all three arrows overlap.
		}
		\label{massdistri}
	\end{center}
\end{figure}

To compare TCGs with other galaxy populations, we also select comparison samples of less actively star-forming galaxies and quiescent galaxies. These comparison samples are selected based on the rest-frame $UVJ$ color cuts \citep{Williams:2009aa} and they do not satisfy the OIRTC selection technique. To ensure like-to-like comparison, ideally we need to select comparison galaxies that are matched to the TCGs in redshift, stellar mass and sample size. However, because the number of quiescent galaxies drop significantly at $z>2$ (\autoref{massz}), we have to slightly reduce the sample size of comparison quiescent galaxies to 2131. Similarly, it is hard to find massive star-forming galaxies that are not TCGs, {as a result the comparison star-forming galaxy sample consists of 2084 galaxies.} The normalized histogram of redshift for all three populations along with the redshift probability distribution ($p(z)$) is plotted in \autoref{zhist}, and the stellar mass distribution is plotted in \autoref{massdistri}. 

\section{Basic properties of TCGs}\label{sec:tcg}
\subsection{850\,$\mu$m flux}
\begin{figure}
	\begin{center}
		\leavevmode
		\includegraphics[scale=0.48]{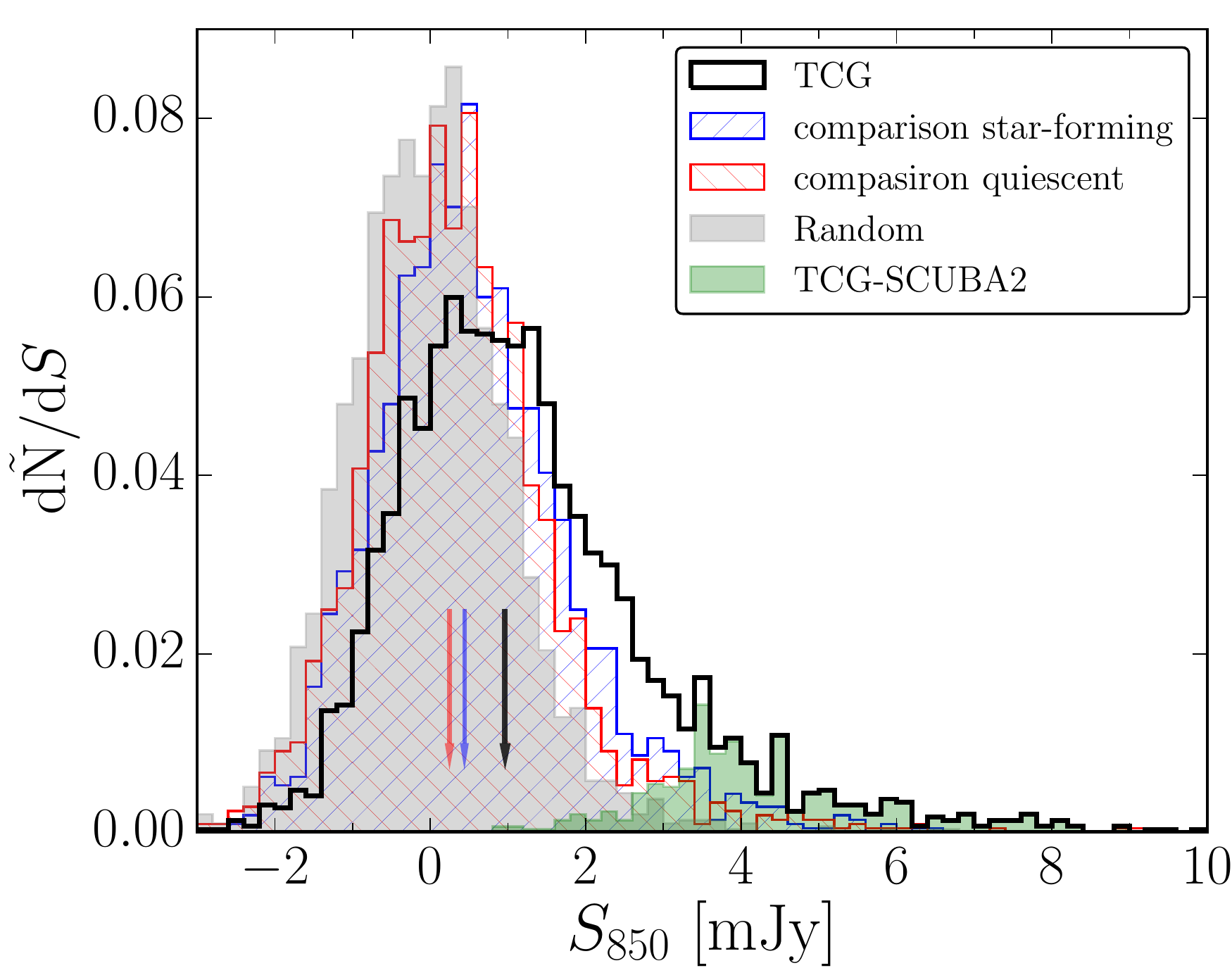}
		\caption{Normalized histogram of 850\,$\mu$m flux for TCGs, comparison quiescent galaxies, and comparison star-forming galaxies. The median values of our sample sources are marked by the vertical bars, with the width representing the bootstrapped uncertainties. The grey regions show the flux distribution of random positions, which is consistent with zero and confirm the detections of all three sample populations. The green regions show the histogram of TCGs that are located close to (within 8$\farcs$7) a SCUBA-2 detection, which as expected all lie in the bright end of the flux distribution. 
		}
		\label{s850hist}
	\end{center}
\end{figure}

We first investigate the 850\,$\mu$m flux distribution based on the SCUBA-2 map that covers all of our sample sources within its footprint (\autoref{overview}). In \autoref{s850hist}, based on the pixel values of the SCUBA-2 map that matched to the locations of our sample sources, we plot the 850\,$\mu$m flux distribution of TCGs, as well as that of the comparison star-forming galaxies and quiescent galaxies. We see a significant offset toward positive values for all three populations compared to the pure noise signals obtained from the pixel values of the random positions. The weighted average for TCGs, comparison star-forming galaxies, and comparison quiescent galaxies are $\langle$S$_{\rm 850}\rangle=$1.25$\pm$0.02, 0.56$\pm$0.02, and 0.36$\pm$0.02\,mJy, respectively, while the median fluxes are S$_{\rm 850, median}=$0.96$\pm$0.04, 0.44$\pm$0.03, and 0.25$\pm$0.03\,mJy, respectively.

We therefore find significant detections at stacked 850\,$\mu$m for all three populations, and the TCGs are the brightest, with a typical 850\,$\mu$m flux of
$\sim$1\,mJy, consistent with them being the dominant population to the 850\,$\mu$m background light (e.g., \citealt{Chen:2013gq}). On the other hand, we also find that there is $\sim$13\% (393/2938) of the TCGs that can be matched to the bright SCUBA-2 detections with S$_{850} \gtrsim 3$\,mJy, within a typical search radius of 8$\farcs$7. Interferometric follow-up observations are needed to confirm their 850\,$\mu$m fluxes, however, as shown in \autoref{sec:clustering} our results are not sensitive to this potential contamination of bright SMGs. 

It is also interesting to see that, if we fit a Gaussian profile only on the fainter half of the flux distribution, the dispersions of both the TCGs and the random noise agree with each other, suggesting that instead of randomly sampling the map flux below the median value (1\,mJy), the fainter TCGs have majority of their 850\,$\mu$m fluxes above 0.4\,mJy, which is the peak of the fit to the fainter half of the flux distribution of TCGs.

On the other hand, of all the 497 SCUBA-2 sources that are located within the mask region of the TCGs (\autoref{overview}), $\sim$60\% (300/497) can be found to have TCGs as their candidate SMG counterpart within the typical search radius of 8$\farcs$7. This demonstrates that the OIRTC selection does not identify all the bright SMGs, and in fact is also incomplete for the faint SMGs as well. Specifically, in \citet{Chen:2016aa} we found that in the training sample  {$\sim$52\% (27/52)} of the ALMA SMGs can be selected through the OIRTC technique, suggesting a completeness of $52\pm12$\% assuming Poisson statistics, a value that is not sensitive to the apparent 850\,$\mu$m flux in the S$_{850} >$1\,mJy regime. 

In the S$_{850} <$1\,mJy regime the completeness could be lower, as tentative evidence of bluer optical--near-IR colors for faint SMGs in this flux regime has recently been claimed by \citet{Hatsukade:2015aa}, which might lead them to fall out of the OIRTC selection. {We stress that, in \citet{Chen:2016aa} there are only three out of 30 ALMA SMGs in the training sample that could be matched to a {\it K}-band source but could not be selected by the OIRTC technique (see Figure 8 in \citealt{Chen:2016aa}). The majority of the ALMA SMGs that are missed by our selection are due to the fact that there is no {\it K}-band counterpart. And because of that, we are not concerned that most of the missing faint SMGs falling into the comparison star-forming galaxy sample. Evidence to support this argument can be found in \autoref{s850hist}. If most of the missing 50\% are part of the comparison star-forming galaxy sample then we should see a similar $S_{850}$ distribution compared with TCGs.

Together with the scenario suggested in \autoref{s850hist} we conclude that, while incomplete,} the majority of 2938 TCGs have 850\,$\mu$m fluxes of S$_{850} \gtrsim 0.4$\,mJy, with a median of S$_{850} \sim 1$\,mJy, representing a first large sample of faint SMG across a $\sim$degree$^2$ field.



\subsection{Relationship with DRGs, EROs, KIEROs, and HIEROs}
Various color cut techniques have been proposed, aiming to select red galaxies in the optical-infrared wavebands that are normally located at $z>1$ and sometimes dusty. The OIRTC technique is unique in that the method targets the most strongly star-forming dust-obscured galaxies and is empirically trained and tested using an ALMA-detected SMG sample at 850\,$\mu$m. The success rate of selecting a $S_{850} > 1$\,mJy SMG among the training sample is $\sim$90\%. Therefore the majority of the TCGs are expected to be dusty. 

While the methods differ, overlaps among populations selected using different techniques are expected. It is therefore informative to compare TCGs to other galaxy populations. Here we consider Distant Red Galaxies (DRGs; \citealt{Franx:2003fr};  $(J-K)_{\rm AB}>1.3$), Extremely Red Objects (EROs; \citealt{Elston:1988zr};  $(R-K)_{\rm AB}>3.6$), EROs with $K$-band and IRAC (KIEROs; \citealt{Wang:2012ys}; $(K-[4.5])_{\rm AB}>1.75$), and EROs with $H$-band and IRAC (HIEROs; \citealt{Wang:2016aa}; $(H-[4.5])_{\rm AB} > 2.25$). 

Based on our UDS parent sample, there are 4582 DRGs, 3889 EROs, 982 KIEROs, and 1003 HIEROs. Thus the number densities of DRGs and EROs are comparable, while they are 30\% more abundant than those of TCGs and a factor of 3-4 times more than those of KIEROs and HIEROs. Among 2938 TCGs, 37\% are DRGs, 13\% are EROs, 19\% are KIEROs, and 25\% are HIEROs. Perhaps more importantly, 41\% of TCGs do not belong to any of the other classifications, demonstrating the uniqueness of the OIRTC selection. 

On the other hand, for each population the fraction that can be selected as TCGs are 24\%, 10\%, 51\%, and 73\% for DRGs, EROs, KIEROs, and HIEROs, respectively. That is, TCGs are the least related to EROs, and majority of the KIEROs and HIEROs meet the OIRTC selection criteria.

\subsection{Rest-frame $UVJ$ color and the selection of quiescent galaxies at $z>1$}\label{subsec:uvj}
Due to a lack of high quality spectra for large samples of the galaxies in particular at $z>1$, color selections in rest-frame $U-V-J$ or $NUV-r^+-J$ utilizing the Balmer/4000\,\AA\,\,breaks have usually been employed to attempt to separate quiescent galaxies from star-forming galaxies (e.g., \citealt{Williams:2009aa,Ilbert:2010aa,Whitaker:2012aa}). Optical/NIR spectroscopic studies have supported the effectiveness of these techniques and show that some quiescent galaxies selected by these techniques indeed appear to have low SFRs (e.g., \citealt{Kriek:2006aa,Whitaker:2013aa}). However, {the influence of dust obscuration} may not be properly accounted for in studies that are only based on optical/NIR data, which by selection is weighted toward less obscured regions. It is therefore crucial to investigate the robustness of these color selections directly in FIR/submillimeter regime. 

\begin{figure}
	\begin{center}
		\leavevmode
		\includegraphics[scale=0.82]{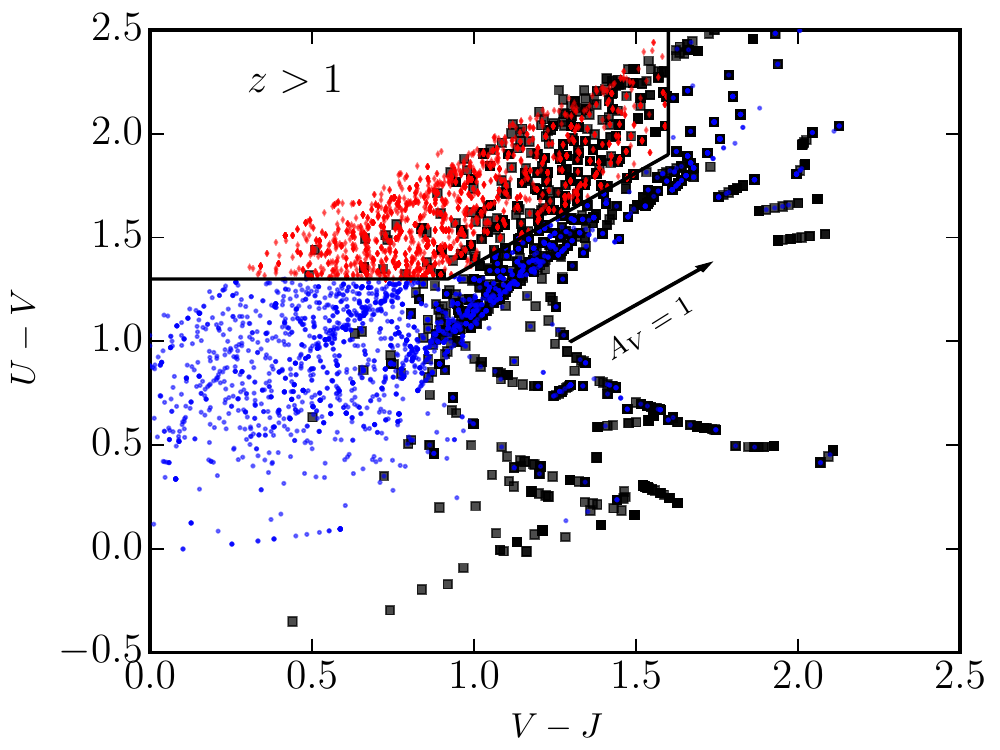}
		\caption{Rest-frame $U-V$ versus $V-J$ diagram for our sample sources at $z>1$. The symbols are the same as \autoref{massz}. Note that the pattern of the distribution is quantized due to the {\sc eazy} template fitting for deriving $z_{\rm photo}$ (\autoref{sec:parent}). The number of TCGs that satisfy the quiescent galaxy color selection is not negligible, suggesting significant contamination from dusty galaxies in the typical quiescent galaxy selections using $UVJ$ color cuts. 
		}
		\label{uvj}
	\end{center}
\end{figure}

\begin{figure}
	\begin{center}
		\leavevmode
		\includegraphics[scale=0.4]{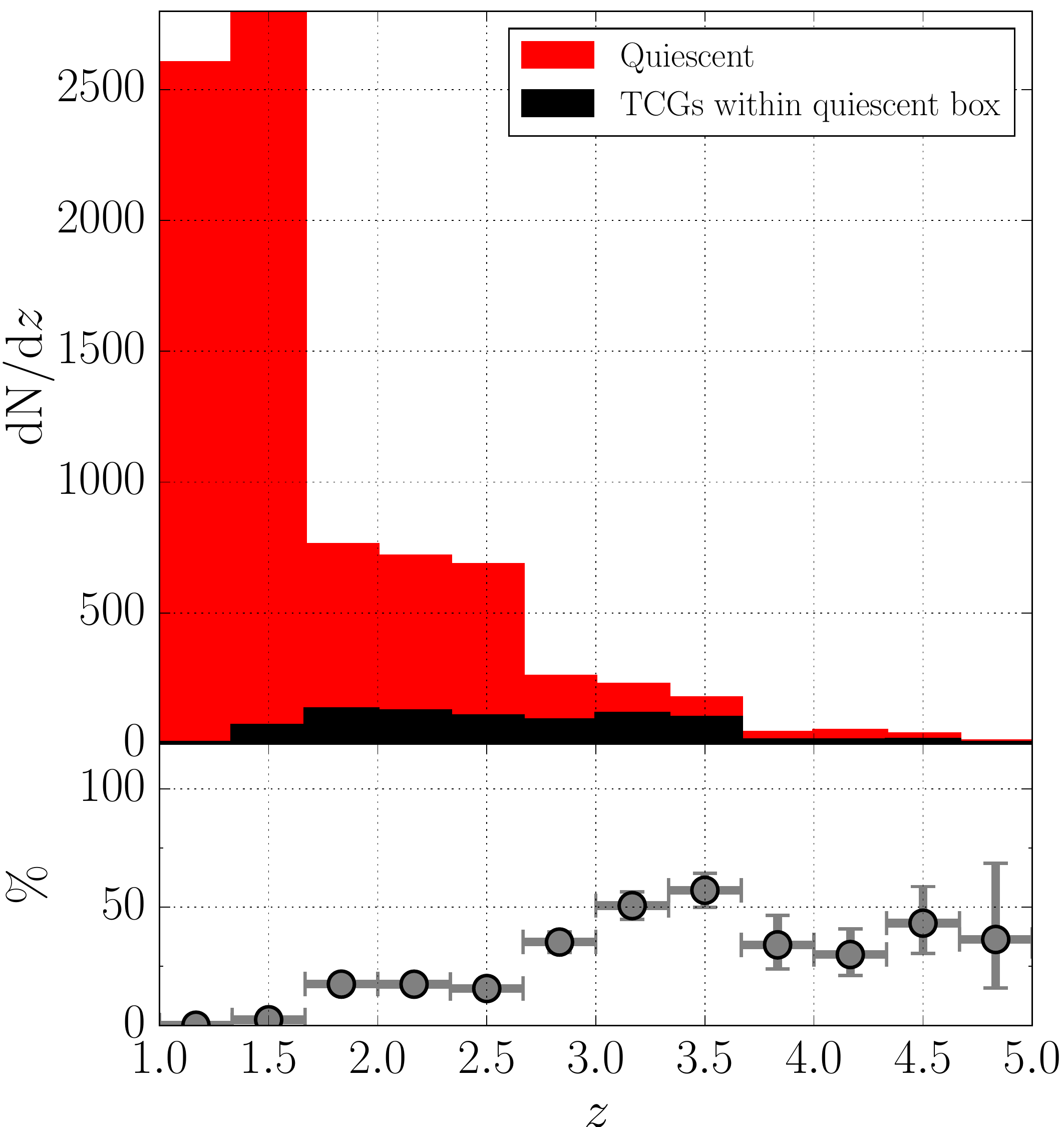}
		\caption{{\it Top}: Histogram of redshift for all quiescent galaxies in the parent sample and the TCGs that are located within the quiescent galaxy color box in the $UVJ$ diagram. {\it Bottom}: The fraction of TCGs in each redshift bin, assuming Poison errors. The contamination fraction is 15-20\% at $z=1.5-2.5$ but rises rapidly to 30-50\% at $z>2.5$, cautioning the use of $UVJ$ colors to select passive galaxies at $z\gtrsim3$.
		}
		\label{tcfract}
	\end{center}
\end{figure}

First in \autoref{uvj} we show the rest-frame $UVJ$ diagram with all our sample sources at $z>1$, showing that the fraction of TCGs, which are very likely obscured dusty galaxies, that are located within the quiescent galaxy selection box is not negligible ($\sim$10\%; 795/8214). The degree of contamination from TCGs is in fact a strong function of redshift. In \autoref{tcfract} we show that the contamination fraction is $15-20$\,\% at $z=1.5-2.5$, and dramatically increases to $30-50$\,\% at $z>2.5$. The quoted values are likely to be underestimations, as some of the dusty galaxies may be missed out from our OIRTC selection. We later highlight in \autoref{sec:discussion} that because of this contamination, the SFRs of the comparison quiescent galaxies at $z>2$ are in fact comparable to the main-sequence star-forming galaxies. {Our finding of the increasing $L_{\rm IR}$ as a function of redshift for the $UVJ$-selected quiescent galaxies is consistent with that of \citet{Viero:2013aa}, whose study was based on the {\it Herschel} data.} We note that the results remain the same at $z<2.5$ if we instead adopt the slightly different {\it UVJ} selection proposed in \citet{Whitaker:2012aa}. 

\begin{figure*}
	\begin{center}
		\leavevmode
		\includegraphics[scale=0.68]{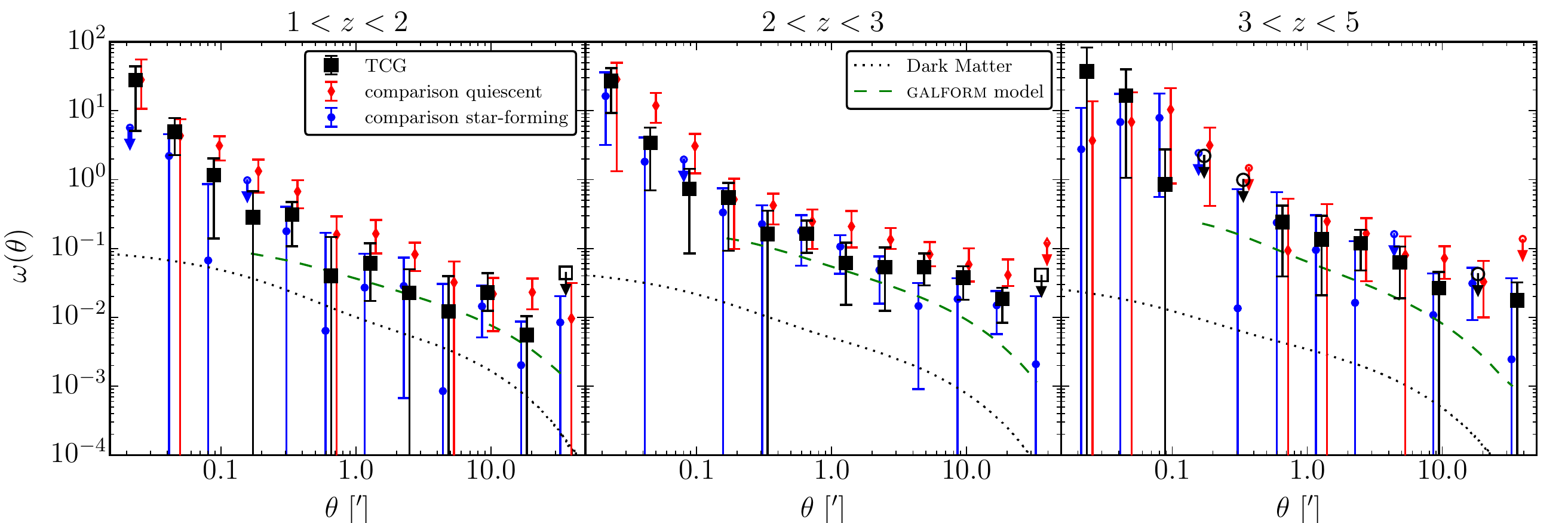}
		\caption{The two-point autocorrelation function of our sample sources at $z=1-5$. For clarity, the data points of the two comparison samples are offset slightly. The dotted curves show the autocorrelation functions of dark matter. We also show in green dashed curves the latest predictions based on the semi-analytic model {\sc galform} \citep{Cowley:2015ab}. At $z<3$, the comparison quiescent galaxies have systematically higher clustering amplitude, whereas TCGs and comparison star-forming galaxies are similarly clustered in all redshift bins.}
		\label{wtheta}
	\end{center}
\end{figure*}

At $z<2$, our results are consistent with previous studies of \citet{Fumagalli:2014aa} and \citet{Man:2014aa}, who both stacked 24\,$\mu$m imaging to attempt to quantify dusty SFRs (Man et al. also stacked the FIR {\it Herschel} imaging at 250, 350, and 500\,$\mu$m) to test the robustness of $UVJ$ diagram in selecting quiescent galaxies. However at $z>2$, our results suggest that the contamination from luminous dusty star-forming galaxies is not negligible, in contrast with the findings of \citet{Fumagalli:2014aa} and \citet{Man:2014aa}. This conclusion is consistent with the radio stacking results in \citet{Man:2014aa}, but they attributed the radio flux enhancement to AGNs.

\section{Clustering and halo mass}\label{sec:clustering}
\subsection{Two point autocorrelation function and large scale power-law fit}
One useful way to study galaxy evolution and the relationship among different populations is to measure the spatial clustering of galaxies, using the two-point autocorrelation functions to estimate the typical mass of their dark matter halos.

Using our sample of $\sim$3000 TCGs, we aim to constrain the clustering strength and infer the host dark matter halo mass of faint SMGs at $z=1-5$ and compare the results to those of our comparison samples.

Following \citet{Chen:2016aa} we calculate the two-point autocorrelation function, $w(\theta)$, using the \citet{Landy:1993aa} estimator.

\begin{equation}\label{eqn:ls}
w(\theta) = \frac{1}{RR}(DD-2DR+RR)
\end{equation}
where DD, DR, and RR are the number of Data-Data, Data-Random, and Random-Random galaxy pair, respectively, counted in bins of angular separation, $\theta$. DR and RR are normalized to have the same total pairs as DD, in a sense that given $N_{S}$ sample sources, $N_R$ random points, $N_{gr}(\theta)$ and $N_{rr}(\theta)$ in the original counts, $DR = [(N_{S}-1)/2N_R]N_{gr}(\theta)$ and $RR = [N_{S}(N_{S}-1)]/N_R(N_R-1)]N_{rr}(\theta)$.

Because our sample sources are located in a single region, $w(\theta)$ needs to be corrected for the integral constraint ({\sc IC}). If the true $w(\theta)$ can be described as a power-law model $w(\theta)_{mod} = A\theta^{-0.8}$ (which has been found to be valid both observationally and theoretically, at the physical separation of $\sim$0.1--10 $h^{-1}$ Mpc), the observed $w(\theta)$ will follow the form 
\begin{equation}\label{eqn:wth}
w(\theta) = w(\theta)_{mod} - IC.
\end{equation}
The integral constraint can be numerically estimated (e.g., \citealt{Infante:1994aa, Adelberger:2005aa}), using the random-random pairs with the following relation: 
\begin{equation}
IC=\frac{\sum_i N_{rr}(\theta_i)w(\theta_i)_{mod}}{\sum_i N_{rr}(\theta_i)}.
\end{equation}

In practice, in \autoref{eqn:ls} we use four times as many random points as the number of sample sources (data points) and repeat the estimate 25 times. Using these 25 estimates we calculate variance, mean $w(\theta)$, as well as mean $N_{rr}$ for the correction of the integral constraint. We then perform $\chi^2$ minimization using \autoref{eqn:wth} to find the best fit $w(\theta)_{mod}$ on $0\farcm2-6'$ scales ($\sim0.2-6 h^{-1}$Mpc), the power-law regime that is shown below. In \autoref{wtheta} we plot the $IC$-corrected $w(\theta)$ of our sample sources, showing that in all three redshift bins the comparison quiescent galaxies have systematically higher clustering amplitude relative to both the TCGs and the comparison star-forming galaxies, and that TCGs and the comparison star-forming galaxies are clustered at a similar level. 

At this stage, however, the error in the amplitude of $w(\theta)_{mod}$ is unrealistically small as the variance only accounts for the shot noise from the sample of the random points and the Poission uncertainties of the $DD$ counts ($DD^{0.5}$). {In addition, the errors are likely correlated between close bins. To estimate the systematic uncertainties due to field-to-field variation as well as to assess the covariance matrix}, we conduct the ``delete one jackknife'' resampling method \citep{Norberg:2009aa}. We first divide the chosen rectangular area\footnote{\label{note3}For the ease of estimating the jackknife uncertainties, we only use sources that are located within a chosen rectangle region with an area of $\sim$0.38 degree$^2$ ($0.61^{\circ}\times 0.63^{\circ}$).}, in which we calculate $w(\theta)$ for the whole sample, into $N_{sub} = 9$ ($3\times$3) equal-size sub-area. Each jackknife sample is defined by discarding, in turn, each of the $N_{sub}$ sub-area into which the whole sample has been split. Each jackknife sample therefore consists $N_{sub}-1$ remaining sub-area, with a volume ($N_{sub}-1$)/$N_{sub}$ times the volume of the full rectangular area. {The covariance matrix is then derived using

\begin{equation}
C_{i,j}=\frac{N_{sub}-1}{N_{sub}}\sum_{k=1}^{N_{sub}}(w(\theta_i)^k-w(\theta_i))(w(\theta_j)^k-w(\theta_j)), 
\end{equation}
where $w(\theta_{i,j})^k$ are the autocorrelation function measured in each jackknife realization. 

We then determine the best fits by minimizing $\chi^2$, which is defined as
\begin{equation}
\chi^2=\sum_i\sum_j(w(\theta_i)^k-w(\theta_i)_{mod})C_{ij}^{-1}(w(\theta_j)^k-w(\theta_j)_{mod}), 
\end{equation}

where $C_{ij}^{-1}$ is the inverse of the covariance matrix $C_{ij}$. The 1\,$\sigma$ errors are estimated based on $\Delta\chi^2=1$.}

To convert the clustering strength to the inferred dark matter (DM) halo mass, $M_{halo}$, we first compute the galaxy bias, $b$, by comparing the $w(\theta)$ of dark matter, $w(\theta)_{DM}$, to our measurement, quantified by the relationship $w(\theta)_{mod}= b^2\times w(\theta)_{DM}$.

To compute $w(\theta)_{DM}$ we first need to obtain the dimensionless dark matter nonlinear power spectrum, $\Delta^2(k)=k^3P(k)/(2\pi^2)$, consisting of one- and two-halo terms. We use the {\sc halofit} code of \citet{Smith:2003aa}, with improved parametrisation provided by \citet{Takahashi:2012aa}, to model $\Delta^2(k)$. We then use Limber's equation to project the power spectrum into the angular autocorrelation function \citep{Limber:1953aa,Peebles:1980aa,Baugh:1993aa}. Specifically we use Equation A6 in \citet{Myers:2007aa} to perform the projection, taking the redshift probability distribution, $p(z)$, of the sample sources into considerations. The example $ w(\theta)_{DM}$ profiles with the $p(z)$ of the $z=1-2$, $z=2-3$, and $z=3-5$ TCGs are shown in \autoref{wtheta}. We fit the DM profile with the power-law $w(\theta)$ to find $A_{DM}$, and then divide our measurements by $A_{DM}$ to obtain $b$. Lastly, we convert $b$ to $M_{halo}$ using the prescription based on the ellipsoidal collapse model of Sheth, Mo \& Tormen (2001). Our results are presented in \autoref{tab1}.

\begin{table*}
	\centering
	\caption{Results of our clustering analyses}
  \begin{threeparttable}
	\begin{tabular}{ccccc}
		\toprule
		Sample  &  $N_S$$^\alpha$  &  $b$	&	$r_0$	&	log$_{10}$(M$_{\rm halo})$    \\		
		  &    &  	&	[$h^{-1}$Mpc]	&	[$h^{-1}$M$_\odot$]   \\
		\midrule
		$z=1-2$	&	&	&	&	\\
		TCGs  			 & 705     & {3.1$^{+0.5}_{-0.6}$}  &	{7.8$^{+1.3}_{-1.5}$}   &  {12.9$^{+0.2}_{-0.3}$}     \\
		comparison star-forming galaxies  &  {523} 	& {2.4$^{+0.6}_{-0.8}$}  & {5.7$^{+1.7}_{-2.2}$}   &  {12.5$^{+0.4}_{-0.9}$}      \\
		comparison quiescent galaxies    &  536 	 & {3.4$^{+0.7}_{-0.8}$}	& {8.5$^{+1.9}_{-2.3}$}	&  {13.1$^{+0.3}_{-0.4}$}  \\
			&	&	&	&	\\
		$z=2-3$	&	&	&	&	\\	
		TCGs  			  &  725 	& {4.0$^{+0.4}_{-0.5}$} & {7.6$^{+0.9}_{-1.0}$}	   &  {12.7$^{+0.1}_{-0.2}$}       \\
		comparison star-forming galaxies  &  {554} 	 & {4.6$^{+0.4}_{-0.4}$} & {8.8$^{+0.8}_{-0.8}$}	&  {12.9$^{+0.1}_{-0.1}$}     \\
		comparison quiescent galaxies    &  550	& {5.7$^{+0.6}_{-0.7}$}	& {11.3$^{+1.3}_{-1.5}$}	&  {13.2$^{+0.1}_{-0.2}$}   \\		
			&	&	&	&	\\
		$z=3-5$	&	&	&	&	\\	
		TCGs  			  &  298 	& {6.9$^{+1.2}_{-1.4}$} & {10.8$^{+2.1}_{-2.5}$}	   &  {12.9$^{+0.2}_{-0.3}$}       \\
		comparison star-forming galaxies  &   {148}	 & {6.1$^{+2.0}_{-3.2}$} & {9.0$^{+3.4}_{-5.1}$}	&  {12.7$^{+0.4}_{-1.3}$}     \\
		comparison quiescent galaxies    &  160 	 & {6.4$^{+1.9}_{-2.8}$}	& {9.8$^{+3.3}_{-4.6}$}	&  {12.7$^{+0.3}_{-0.9}$}   \\				
		\bottomrule
	\end{tabular}
	\label{tab1}
    \begin{tablenotes}
    	\item $^\alpha$ For the ease of estimating the jackknife uncertainties, we only use sources that are located within a chosen rectangle region with an area of $\sim$0.38 degree$^2$ ($0.61^{\circ}\times 0.63^{\circ}$).
    \end{tablenotes}
  \end{threeparttable}
 \end{table*}

\begin{figure}
	\begin{center}
		\leavevmode
		\includegraphics[scale=0.87]{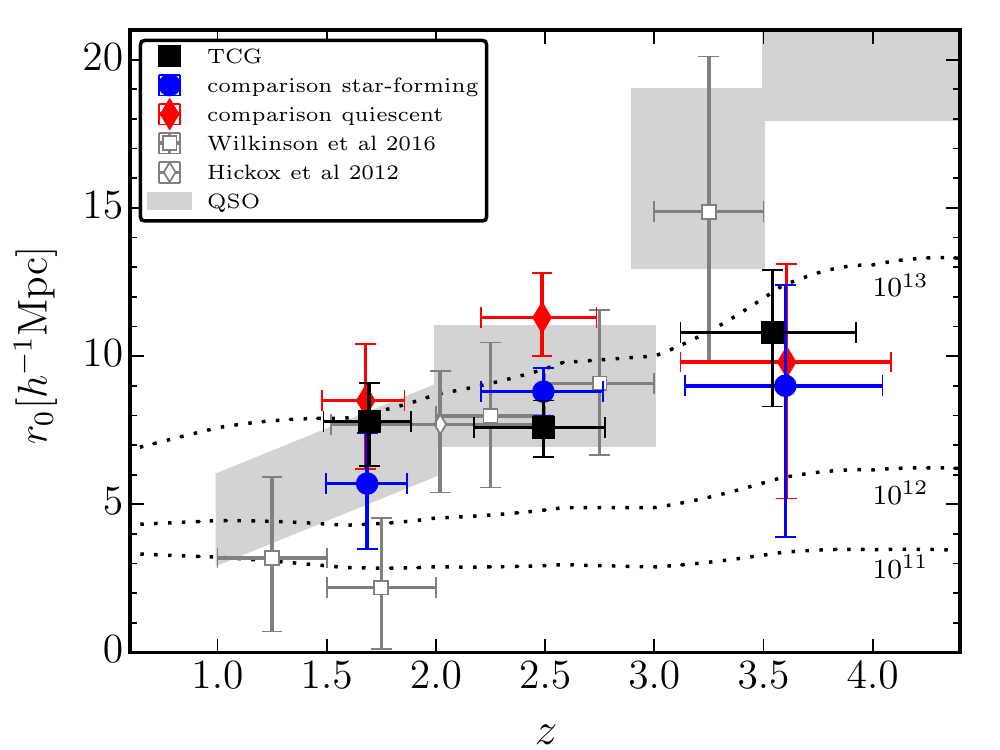}
		\caption{Autocorrelation length $r_0$ as a function of redshift. Data points of our measurements are plotted. We also plot the measurements of bright SMGs adopted from the literature presented in \citet{Hickox:2012kk,Wilkinson:2016aa}. The grey boxes outline the regions of quasars \citep{Myers:2006aa,Porciani:2006aa,Shen:2007aa,Eftekharzadeh:2015aa}. Dotted curves roughly outline the corresponding $r_0$ values for DM halos of different masses. Our results suggest that faint SMGs have similar DM halo masses to the comparison star-forming galaxy sample, but are residing in the halos that have lower masses compared to the comparison quiescent galaxy sample.}
		\label{r0}
	\end{center}
\end{figure}

In addition, in the case of small angular separations ($\theta \ll 1$\,rad) and assuming no clustering evolution in redshift, we can also derive the autocorrelation length $r_0$ by inverting Limber's equation (e.g., \citealt{Peebles:1980aa,Myers:2006aa,Hickox:2011aa}) as the following

\begin{equation}
A = H_\gamma\frac{\int_0^\infty (dN/dz)^2E_z\chi^{1-\gamma}dz}{[(dN/dz)dz]^2}r_0^\gamma, 
\end{equation}
where $H_\gamma = \Gamma(0.5)\Gamma(0.5[\gamma-1])/\Gamma(0.5\gamma)$, $\gamma = 1.8$, $E_z = H_z/c$ ($H_z=H_0\sqrt{\Omega_m(1+z)^3+\Omega_\lambda}$), $\chi$ is the radial comoving distance, and $dN/dz$ is the redshift probability distribution $p(z)$. Similar to the autocorrelation function, the advantage of $r_0$ is that its derivation does not involve any assumption about the dark matter halo model, which makes $r_0$ a particularly useful quantity to compare with the model predictions. The results are also presented in \autoref{tab1}, and plotted in \autoref{r0}. 

\section{Discussion}\label{sec:discussion}
As shown in \autoref{wtheta} and \autoref{r0}, in general we find that, at $z=1-3$ the dusty galaxies (TCGs) are similarly clustered, and therefore reside in similar mass halos, as the less dusty comparison star-forming galaxies, while the comparison quiescent galaxies are significantly more clustered compared to the other two populations. The halo mass that separates the comparison quiescent galaxies and TCGs/comparison star-forming galaxies appears to be at log$_{10}$(M$_{\rm halo}$) $\sim13.0$\,$h^{-1}$M$_\odot$ regardless of redshift. The separation disappears at $z>3$, as discussed in \autoref{sec:5p2} this could be due to the increasing contamination of dusty galaxies on the {\it UVJ}-selected quiescent populations. The latest model predictions by \citet{Cowley:2015ab} on the clustering of both faint and bright SMGs are systematically lower in all three redshift bins. 

\subsection{Halo quenching}
The general picture of quiescent galaxies having a stronger clustering strength compared to the stellar mass-matched star-forming galaxies is in agreement with what was found recently in \citet{Hartley:2013aa}, \citet{Sato:2014aa}, \citet{McCracken:2015aa} and \citet{Lin:2016aa}, but contrary to what was reported in \citet{Bethermin:2014aa}. This finding implies that halo mass could play a major role in quenching the star formation, in line with theoretical models that advocate the halo quenching scenario (e.g., \citealt{Cen:2011aa}). These models have suggested that gas heating, which could be attributed to virial shocks, AGN feedback, and/or gravitational energy of cosmological accretion, in halos above 10$^{12}$\,M$_\odot$ prevents sufficient gas cooling and therefore inhibits the continued formation of stars (e.g., \citealt{Birnboim:2003aa,Bower:2006aa,Dekel:2008aa}). Recent hydrodynamical simulations run by \citet{Gabor:2015aa} demonstrated that the hot halo quenching can be used to explain two distinct quenching mechanisms, mass quenching and environmental quenching, that are proposed from observations by \citet{Peng:2010qy}.

\begin{figure*}
	\begin{center}
		\leavevmode
		\includegraphics[scale=0.67]{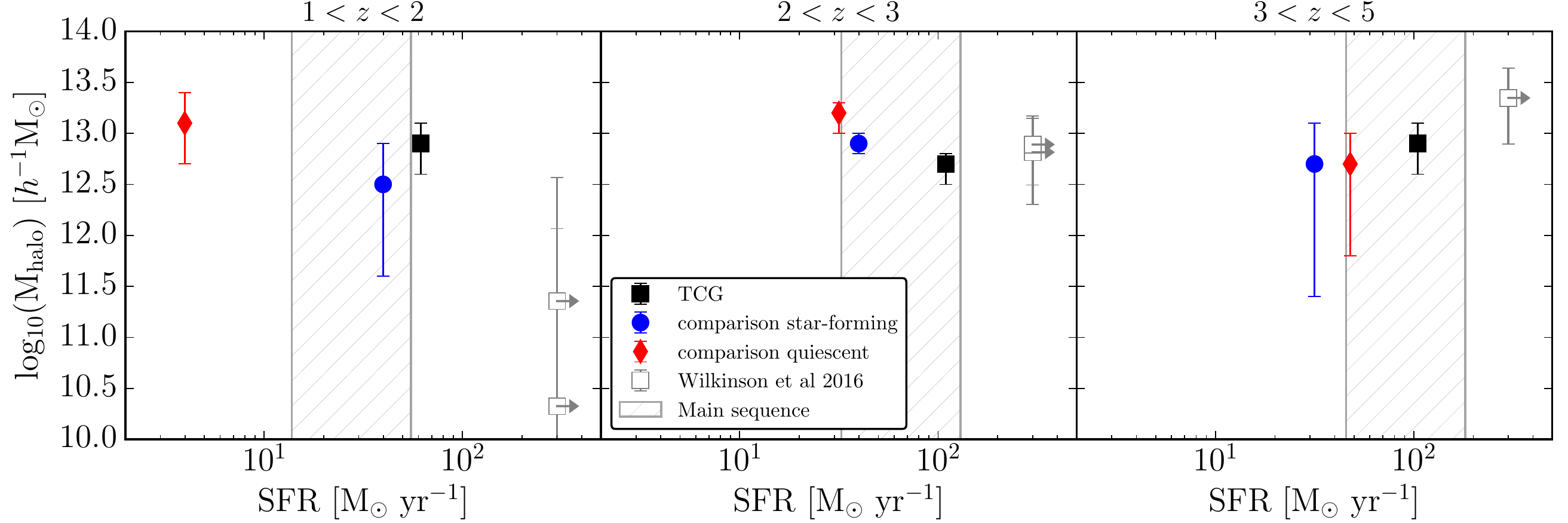}
		\caption{Halo mass versus star formation rate in three redshift bins. We mark the measurements of bright SMGs from \citet{Wilkinson:2016aa}, with the star formation rate limit estimated based on their SCUBA-2 survey depth. The grey vertical bands mark the SFR--M$_\ast$ main-sequence based on \citet{Speagle:2014aa} and median stellar masses/redshifts of the sample galaxies. A lack of correlation is observed among star-forming samples, suggesting that the cause of increased SFR could lie in their local galactic environments.}		
		\label{hmass}
	\end{center}
\end{figure*}

\subsection{A lack of correlation between halo mass and star formation rate}\label{sec:5p2}
 The lack of dependency on the clustering strength between TCGs and the comparison star-forming galaxies, despite the fact that both populations differ significantly in 850\,$\mu$m flux, is perhaps surprising. {It is possible that due to the incompleteness of our selection, some faint SMGs are missed by the OIRTC technique and instead fall into the comparison star-forming sample. However as shown in \autoref{s850hist} and later in this section, the fraction of those missed faint SMGs is likely to be low, given that on average the comparison star-forming galaxies have much lower $S_{850}$ and SFR. We therefore argue that the similarity of the clustering strength between TCGs and the comparison star-forming galaxies is not caused by the incompleteness of our selection.}

To compare to the galaxies that have even higher 850\,$\mu$m fluxes, in \autoref{r0} we also show the clustering measurements of bright SMGs from the literature. As a result of their low space density, the measurements of the bright SMGs clustering have suffered from small number statistics. The most constrained measurements by far have been made using a cross-correlation technique \citep{Hickox:2012kk,Wilkinson:2016aa}, and as shown in \autoref{r0} the results are consistent with those of TCGs and the comparison star-forming galaxies. 

If the clustering measurements of TCGs are representative of the overall faint SMG population, meaning both bright and faint SMGs reside in similar mass halos, we can roughly estimate the typical lifetime of faint SMGs based on their number counts. Since recently both observations and theoretical models have suggested little variations of redshift distributions between bright and faint SMGs (e.g., \citealt{Simpson:2014aa,Cowley:2015aa,Bethermin:2015aa}), together with the fact that faint SMGs with $S_{850} \sim 1$\,mJy are about 5-10 times more abundant than bright SMGs, we expect the lifetime of faint SMGs to be longer than that of bright SMGs by a similar factor. Based on the clustering measurements and the estimated gas depletion timescale, the typical lifetime of bright SMGs is $\sim$100\,Myr (e.g., \citealt{Hickox:2012kk,Bothwell:2013lp}). The lifetime of faint SMGs is therefore expected to be $\sim$0.5-1\,Gyr. As shown below, unlike bright SMGs, faint SMGs are likely located on the extragalactic star-forming main sequence. Recent ISM studies have suggested that the main-sequence galaxies at $z\sim2$ have a gas depletion timescale of $\sim$0.5-1\,Gyr (e.g., \citealt{Genzel:2015aa,Scoville:2015aa}), which is consistent with our estimation.

To put our sample sources into a more general context, we estimate the SFR of our samples through the FIR SED fitting, using the data from both SPIRE at 250, 350, and 500\,$\mu$m and SCUBA-2 at 850\,$\mu$m. We fit the median stacked SED of all three populations, split into three redshift bins, using the templates provided by \citet{Magdis:2012aa}, which are suitable for IR main-sequence galaxies. We then derive the SFR based on the best fit template SED and the Kennicutt star formation law assuming Chabrier initial mass function (i.e., SFR = L$_{\rm IR}$/10$^{10}$L$_\odot$; \citealt{Kennicutt:1998p5718}). We plot the results against halo mass in \autoref{hmass}, confirming the scenario suggested based on the 850\,$\mu$m flux and showing that, at the same redshift bin, the clustering strength does not depend on SFR, though the quiescent galaxies at $z<3$ are significantly more clustered.

Lastly, it is also worth pointing out that the problem of selecting quiescent galaxies at $z>2$ using the $UVJ$ diagram, which is highlighted in \autoref{subsec:uvj}, manifests itself in \autoref{hmass}. We find that the SFR of the comparison quiescent galaxies is comparable to that of the comparison star-forming galaxies at $z>2$, both of which lie at the lower edge of the main sequence of star-forming galaxies \citep{Speagle:2014aa}, suggesting that the contamination from dusty galaxies (less dusty than the TCGs, so missed by the OIRTC selection) is not negligible and affects the statistics of the quiescent galaxy population selected through the $UVJ$ diagram at $z>2$. The increasing contamination from dusty galaxies at higher redshifts could lead to underestimations of the clustering on truly quiescent populations at $z>2$, and could explain the consistent clustering strength that we observe between TCGs and the comparison quiescent galaxies at $z=3-5$, although it could also simply due to larger uncertainties caused by smaller sample size.

\subsection{What drives the enhancement of star formation in TCGs?}
The intriguing finding of TCGs, less FIR luminous star-forming galaxies, and bright SMGs all residing in similar halos prompts the question of what differentiates these systems? While star formation is fed by gas, the cosmic accretion rate of pristine gas for all three populations are expected to be similar given similar halo masses, suggesting that the difference could lie in their local galactic environments. Recent studies of the interstellar medium, as traced by CO or dust, in galaxies on and/or above the main sequence at $z\sim2$ have found that at a fixed stellar mass, galaxies with higher SFRs tend to have more molecular gas mass \citep{Bothwell:2013lp,Tacconi:2013aa,Scoville:2015aa}. However, it is claimed that the increase in gas mass does not fully account for the increase in SFR, suggesting more efficient star formation and therefore shorter gas depletion times (e.g., \citealt{Genzel:2015aa}). This can alternatively be explained by the fact that the slope of Kennicutt-Schimit law is $>1$ \citep{Kennicutt:1998p5718}.

\begin{figure}
	\begin{center}
		\leavevmode
		\includegraphics[scale=0.8]{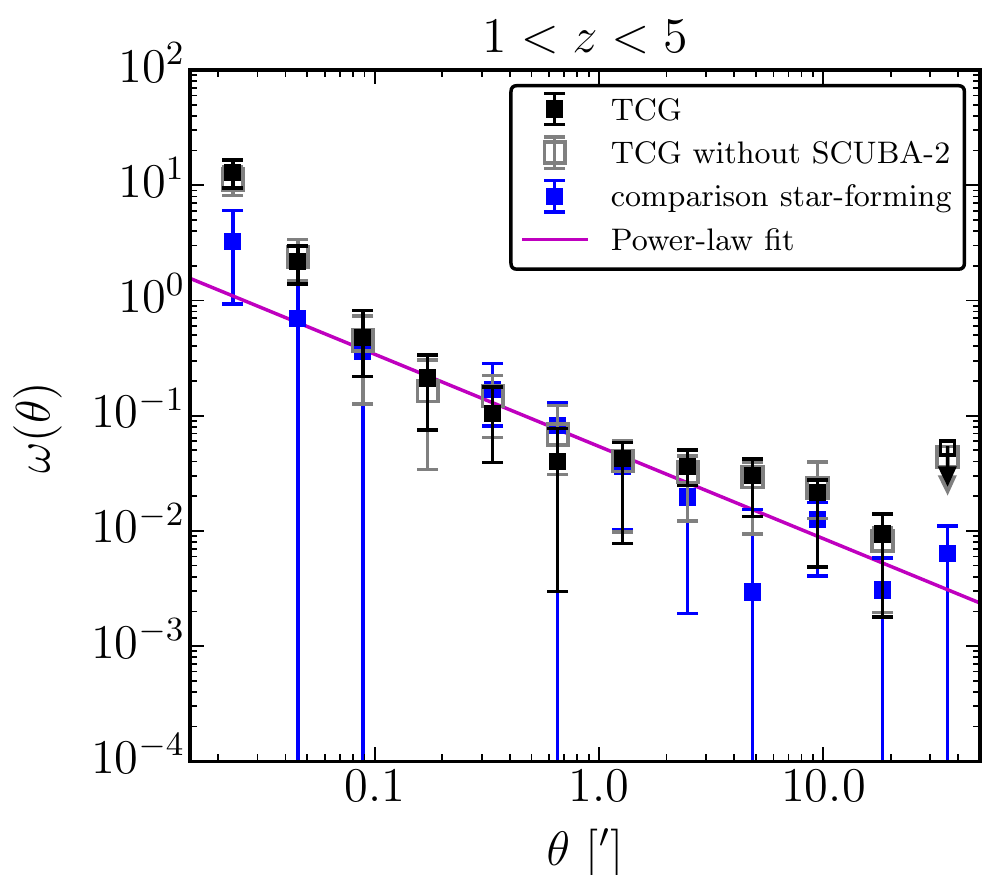}
		\caption{Same as \autoref{wtheta} but with a redshift bin of $z=1-5$. We also plot the measurements of those TCGs that are not in close proximity to any SCUBA-2 detections, which does not show a significant change. A tentative enhancement of the autocorrelation function at the smallest separation bin compared to the comparison star-forming galaxy suggest some physically associated pairs among faint SMGs.}
		\label{wtheta_zgt1lt3}
	\end{center}
\end{figure}

Locally, it has been shown that galaxies undergoing interactions and/or dynamical instability (e.g., bars) can increase their star-formation efficiency  \citep{Saintonge:2012aa}. At $z\sim2$, owing to modest spatial resolution and sensitivity, the morphological separation between interacting galaxies and isolated disk becomes ambiguous. Statistical studies of pair separations, such as two-point autocorrelation function, could provide some evidence of physically associated close pairs. In \autoref{wtheta_zgt1lt3} we again show the autocorrelation function of our sample TCGs and comparison star-forming galaxies in the redshift range of $z=1-5$. Despite both agree on large scales, we find a tentative excess in the smallest separation bin at $\theta=1''-2''$ ($\sim$8--17\,kpc) for the TCGs compared to the comparison star-forming galaxies. This result remains unchanged if we remove those TCGs that associate with bright SMGs, as most studies have suggested that bright SMGs are mergers (e.g., \citealt{Alaghband-Zadeh:2012aa,Chen:2015aa}). Further investigations with spectroscopic follow-up observations are needed but if true our results suggest that at least some fraction of faint SMGs may be mergers or physically associated pairs undergoing interactions, and the enhancement of dynamical perturbations may contribute to the increase in star formation rate.

\section{summary}\label{sec:sum}
Based on a newly developed technique that tuned to efficiently select SMGs combining three optical--near-infrared colors ($z-K$, $K-$[3.6], [3.6]-[4.5]), we apply this color selection technique, OIRTC, to a $K$-band based parent sample with multi-wavelengths photometry in the UKIDSS-UDS field. We have identified 2938 OIRTC-selected galaxies, dubbed TCGs, and conducted analyses to assess their basic properties in terms of 850\,$\mu$m flux, SFR, redshift, stellar mass, and their relationship to other galaxy populations. Exploiting the degree$^2$-scale area of UKIDSS-UDS we measure the spatial clustering of TCGs and derive their halo masses. We also select comparison quiescent galaxy and less FIR luminous star-forming galaxy samples that are matched in redshift and stellar mass to the TCGs to compare their properties. Our findings are summarized in the following:

\begin{enumerate}
	\item We find that the redshift distribution of TCGs spans $z=1-5$, with a median redshift of $z=2.23\pm0.02$. They have a median stellar mass of log$_{10}$(M$_\ast$/M$_\odot$) $=10.51\pm0.02$, and a median 850\,$\mu$m flux of S$_{850} = 0.96\pm0.04$\,mJy, implying SFR$\sim$100\,M$_\odot$yr$^{-1}$. Based on the 850\,$\mu$m flux distribution we argue that majority of TCGs have S$_{850} > 0.4$\,mJy (\autoref{s850hist}). Therefore the OIRTC selection, while incomplete, offers a new route to select large samples of faint SMGs that are historically hard to detect in the single-dish FIR/submillimeter surveys due to confusion.
	
	
	\item We find a non-negligible fraction of TCGs that are located in the quiescent galaxy color selection box in the $UVJ$ diagram, and the fraction of contamination increases with redshift. At $z<1.5$ the contamination fraction is close to zero, while at $1.5<z<2.5$ the fraction is 15-20\% and at $z>2.5$ it is 30-60\% (\autoref{tcfract}). We find similar results when we examine the SFR of the comparison quiescent galaxy sample. Although with TCGs excluded, at $z>2$ the SFR of the comparison quiescent galaxies are similar to the comparison star-forming galaxies, both lie on the SFR-M$_\ast$ main sequence, while at $z<2$ the comparison quiescent galaxies are indeed quiescent with on average specific SFR significantly below the main sequence.

		
	\item The two-point autocorrelation functions suggest that TCGs have a typical halo mass of log$_{10}$(M$_{\rm halo}$) $=12.9^{+0.2}_{-0.3}$, $12.7^{+0.1}_{-0.2}$, and $12.9^{+0.2}_{-0.3}$\,$h^{-1}$M$_\odot$ at $z=1-2$, $2-3$, and $3-5$, respectively. These values are consistent with those of bright SMGs and the comparison star-forming galaxies (\autoref{r0}). If the clustering results of TCGs are representative to the faint SMG population as a whole, our results suggest that the halo mass in which most SMGs reside is independent of 850\,$\mu$m flux and star-formation rate. In addition, based on the number counts and the derived halo mass, we estimate a typical faint SMG lifetime of 0.5--1\,Gyr, compared to $\sim$0.1\,Gyr for bright SMGs \citep{Hickox:2012kk}.
	
	\item We find tentative evidence that TCGs have an enhancement on $\sim8-17$\,kpc scales in their autocorrelation function compared to the comparison star-forming galaxies, despite their autocorrelation functions agree on the larger scales, suggesting that some of the faint SMGs are physically associated (\autoref{wtheta_zgt1lt3}) on these scales, perhaps reflecting a merging origin in their triggering.

\end{enumerate}

\section{Acknowledgments}
We acknowledge the referee for a helpful report that has improved the manuscript. C.-C.C., I.R.S. acknowledge support from the ERC Advanced Investigator programme DUSTYGAL 321334. I.R.S. also acknowledges support from a Royal Society/Wolfson Merit Award and STFC through grant number ST/L00075X/1. A.M.S. acknowledges financial support from an STFC Advanced Fellowship (ST/H005234/1) and the Leverhulme Foundation.
This work was performed in part at the Aspen Center for Physics, which is supported by National Science Foundation grant PHY-1066293.
This research made use of Astropy, a community-developed core Python package for Astronomy \citep{Astropy-Collaboration:2013aa}. This research has made use of NASA's Astrophysics Data System. The authors wish to recognize and acknowledge the very significant cultural role and reverence that the summit of Mauna Kea has always had within the indigenous Hawaiian community. We are most fortunate to have the opportunity to conduct observations from this mountain.

\end{CJK}
\bibliography{bib}

\end{document}